\documentclass[aps,prb,reprint,superscriptaddress]{revtex4-1}

\usepackage{graphicx}

\begin{document}

\title{Vertical Field-Effect Transistor Based on Wavefunction Extension}

\author{A. Sciambi}
\affiliation{Department of Applied Physics, Stanford University, Stanford CA 94305-4045 USA}
\affiliation{SIMES, SLAC National Accelerator Laboratory, 2575 Sand Hill Road, Menlo Park, CA 94025 USA}
\author{M. Pelliccione}
\affiliation{Department of Applied Physics, Stanford University, Stanford CA 94305-4045 USA}
\affiliation{SIMES, SLAC National Accelerator Laboratory, 2575 Sand Hill Road, Menlo Park, CA 94025 USA}
\author{M. P. Lilly}
\affiliation{Center for Integrated Nanotechnologies, Sandia National Laboratories, Albuquerque, New Mexico 87185, USA}
\author{S. R. Bank}
\affiliation{Materials Department, University of California Santa Barbara, Santa Barbara CA 93106 USA}
\affiliation{Electrical and Computer Engineering Department, University of Texas at Austin, Austin TX 78758 USA}
\author{A. C. Gossard}
\affiliation{Materials Department, University of California Santa Barbara, Santa Barbara CA 93106 USA}
\author{L. N. Pfeiffer}
\affiliation{Department of Electrical Engineering, Princeton University, Princeton NJ 08544 USA}
\author{K. W. West}
\affiliation{Department of Electrical Engineering, Princeton University, Princeton NJ 08544 USA}
\author{D. Goldhaber-Gordon}
\affiliation{Department of Physics, Stanford University, Stanford CA 94305-4045 USA}
\affiliation{SIMES, SLAC National Accelerator Laboratory, 2575 Sand Hill Road, Menlo Park, CA 94025 USA}

\date{\today}

\begin{abstract}
We demonstrate a mechanism for a dual layer, vertical field-effect transistor, in which nearly-depleting one layer will extend its wavefunction to overlap the other layer and increase tunnel current. We characterize this effect in a specially designed GaAs/AlGaAs device, observing a tunnel current increase of two orders of magnitude at cryogenic temperatures, and we suggest extrapolations of the design to other material systems such as graphene.
\end{abstract}

\maketitle
Quantum transistors, those that rely on quantum mechanical transport processes for operation, have become an important research direction as conventional transistors are hindered by the emergence of those same effects at the nanoscale.\cite{ITRS2009} In this vein, we present here a novel mechanism for a vertical field-effect transistor, wherein the adjustable subband energy of a planar quantum well modifies the vertical extent and overlap of its bound wavefunction with another parallel well. Unlike past quantum transistors that utilize tunnel resonances of aligned subbands\cite{Kolagunta, Simmons, Khurgin} or single electron levels in quantum dots,\cite{QD} this simple design is not sensitive to lateral dimensions and should be operable down to the few-nanometer scale in suitable materials. We call the resulting device the Wavefunction Extension Transistor (WET).

\begin{figure}[b]
\includegraphics[width=7.5cm]{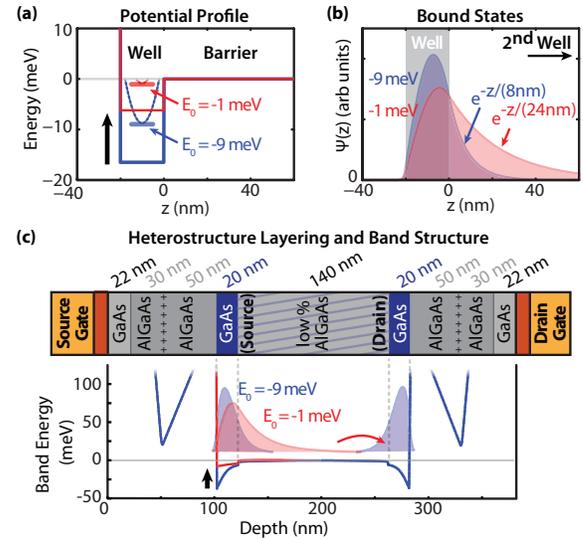} 
\caption{(a) Potential profile of a shallow barrier next to a rectangular well that is raised so the energy $E_0$ of its bound subband edge goes from -9 meV to -1 meV. (b) Simulated wavefunctions corresponding to the two subband energies, with the raised subband having a greatly extended wavefunction. (c) WET composition, consisting of a GaAs/AlGaAs heterostructure with gates and dielectric on either side. A self-consistent simulation of the conduction band edge for the structure is shown below,\cite{Snider} with the source gate varying the subband edge of the source quantum well from -9 meV to -1 meV and extending its associated wavefunction.}
\label{Fig1}
\end{figure}

Within a quantum well containing a single subband, the out-of-plane momentum and characteristic length scale for barrier penetration for bound electrons is solely determined by the height of the barrier above the bottom of the subband, together with the effective mass, a property of the quantum well material. In a WET, this barrier normally inhibits tunneling, but its height can be reduced by electrostatically raising the well containing the subband (Fig 1a). The rate of exponential decay of the subband wavefunction into the barrier scales roughly as the square root of the effective barrier height, so reducing the height nearly to zero causes a subband wavefunction to greatly extend toward the opposing well (Fig 1b). Such spreading leads to an increase in wavefunction overlap and tunneling. Substantial tuning of wavefunction extension and overlap is enabled by (1) having a wide potential barrier separating the two wells to maximize the effect of wavefunction decay, and (2) having the barrier height as low as possible to maximize current while ensuring energy levels in the two wells can be separately manipulated. This need for a low and wide barrier means wavefunction extension has not been observed in more conventional bilayer quantum well systems with high, narrow barriers,\cite{Simmons, Eisenstein2} strongly-coupled wells,\cite{Manoharan} or high interlayer biases.\cite{THETA} Modulating wavefunction overlap has been proposed before for a field-effect tunnel transistor,\cite{Khurgin} but in the context of laterally shaping wavefunctions using multiple side gates.

In this paper, we simulate and experimentally characterize a proof-of-principle transistor designed according to this scheme: we measure at 4.2K the tunneling between source and drain layers epitaxially grown in a GaAs/AlGaAs heterostructure (Fig 1c, top), as a function of voltages on top and back gates. In this type of structure, simply bringing a layer near depletion using surface gates should dramatically increase wavefunction overlap (Fig 1c, bottom) and the related vertical tunnel conductance, as indeed we observe empirically. This puts the WET into a very small subset of transistors\cite{Simmons} where the conductance is tuned using a gate located outside the channel region, with the source or drain intervening between gate and channel. After characterizing the GaAs/AlGaAs device, we conclude by proposing that a WET operating at room temperature with improved switching characteristics could be constructed based on two parallel layers of graphene. We note here that both the GaAs device and the proposed graphene device are orders of magnitude from the current densities and on-off ratios of commercial transistors.\cite{Baker} That said, we believe that this novel mechanism of current-modulation might be useful as new materials and fabrication techniques arise.

In order to optimize the tunable tunneling in our GaAs/AlGaAs device, we use self-consistent one-dimensional Schr\"odinger-Poisson solvers\cite{Snider} to guide our design of a pair of GaAs/AlGaAs quantum wells separated by a wide, low-energy barrier. We estimate tunnel rates using Bardeen's formalism,\cite{Bardeen} which calculates the overlap of wavefunctions constrained to opposing sides of the barrier. This yields a tunneling matrix element between wells of equal subband energy that scales roughly as $T(E_0) \propto \sqrt{|E_0|} e^{-w\sqrt{2m|E_0|}/\hbar}$, where $E_0$ is the (negative) energy of the subband edge relative to the barrier, $w$ is the width of the wide barrier, and $m$ is the effective mass of an electron in the barrier.\cite{Clerc} 

\begin{figure}[b]
\includegraphics[width=7.5cm]{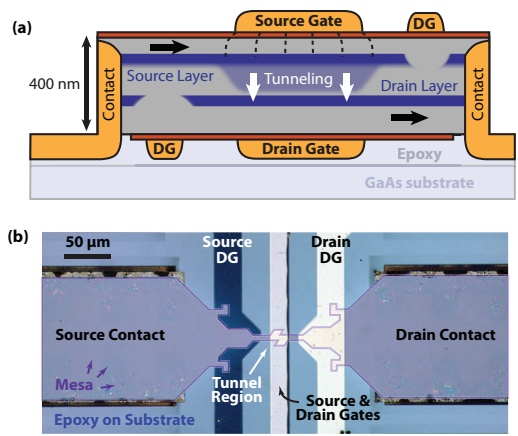} 
\caption{(a) Schematic of WET device (not to scale), with depletion gates (DG) limiting access of the contacts to the tunneling region. (b) Photograph of the finished sample with the GaAs mesa, outlined in purple, supported by epoxy. Hook-shaped protrusions in the mesa are visible where gates overlap, included to ensure continuity during gate deposition over the mesa step.}
\label{Fig2}
\end{figure}

Following optimization, we focus on the structure shown in Fig. 1c, which we have labeled H1 and use for all measurements unless otherwise noted. The structure contains a pair of two-dimensional (2D) electron layers each residing in a 20 nm-wide GaAs well, and separated by a 140 nm-wide, Al$_{0.02}$Ga$_{0.98}$As barrier. The bilayer system is sandwiched between 80 nm Al$_{0.3}$Ga$_{0.7}$As spacers that are delta-doped near their midpoint, and then between 22 nm GaAs caps. After growth in an MBE system,\cite{growers} this heterostructure was found to have source and drain layer densities of 3.1 and 2.9 $\times 10^{11}$ cm$^{-2}$ with mobilities of 3 and 1$\times 10^6$ cm$^2$/Vs, respectively, all at 4.2K. The densities are within 10\% of their simulated values.

The 2.0\% Al barrier, measured precisely during growth with reflection high-energy electron diffraction (RHEED), was found empirically to be the ideal balance between large barrier modulation and large tunnel current. For a 1.0\% barrier, we observe that the two wells are not decoupled and for a 3.0\% barrier, the minute tunneling is difficult to measure. Determining this optimal percentage from first principles is difficult since the barrier position relative to the Fermi energy can vary by more than 10 meV depending on the exact well shape. Our simulation predicts that the 2.0\% barrier is too high (7 meV above $E_F$) to measure the strong gate modulation of tunnel conductance described later. With this disparity in mind, we have adjusted the Al concentration to 0.5\% in simulation, and we use this value to demonstrate wavefunction extension in Fig 1c. This shallower barrier is predicted to be around 1 meV above $E_F$.

To determine tunneling between layers, we employ front and back depletion gates\cite{Eisenstein} (DGs, Fig 2a) to limit access of the sample contacts to only one layer each. In this way, modulation of tunneling by the source and drain gates can be measured by simply applying a bias between the contacts. Backside lithography is accomplished by first fabricating the front side of a sample chip with mesas, contacts, alumina gate dielectric, and gates, and then epoxying that chip face down to a second GaAs substrate. The original substrate is mechanically thinned to 30 $\mu$m, and then chemically etched to 400 nm using selective removal of etch-stop layers grown into heterostructure.\cite{EBASE} After etching, only the mesa remains of the original substrate, with bare epoxy supporting it as well as features off the mesa like the frontside gates and subsequently-added backside gates (Fig 2b). With the flipped sample in mind, we refer to the original backside quantum well as the source and the original frontside well as the drain. The source and drain gates areas are 200 $\mu$m$^2$, allowing accurate measurements of tunnel modulation between source and drain by excluding background tunneling or tunneling induced by fringe fields.

The measurement itself is performed with the application of a 100 $\mu$V AC excitation driven at 152 Hz between the source and drain layers while a lock-in measures the differential tunneling conductance. We confirmed that the bias between the two layers was set by this AC excitation: the gate dielectric ensures that the measured AC current was always many orders of magnitude larger than any DC gate leakage current. Additionally, we find that interlayer biasing up to tens of mV minimally impacts the densities of the layers, by less than 5 $\times$ 10$^8$ cm$^{-2}$/mV as determined from Shubnikov-de Haas oscillations.

As expected, tunneling measurements as a function of source and drain gate voltages reveal a pair of conductance ridges (Fig 3a, dotted lines) at 4.2K associated with near-depletion in the respective layers. Approaching these ridges from the high-density side, the tunnel conductance increase confirms the predicted wavefunction extension. These ridges differ from the resonances often seen in tunneling between low-dimensional systems where their energy-momentum dispersions coincide.\cite{Eisenstein2} Such resonances are absent here (Fig 3a, gray line where densities are matched), possibly due to elastic scattering in the interlayer that creates momentum transfer.

For comparison, we show a density-matched tunneling resonance from a similar structure (H2) with a narrower 70 nm barrier (Fig 3b -- this structure was thinned only to 10 $\mu$m and so required proportionally larger source gate voltages.) In this heterostructure, the near-depletion ridges are also apparent, though mostly obscured by density-matched resonant tunneling (gray line). In a regime lacking interlayer scattering, the WET could operate without impact from energy-momentum constraints if the two layers were set to equal densities and then depleted simultaneously. On Fig 2b, this represents moving from the upper right corner along the gray line towards the tunnel maximum near the plot center. Alternatively, the constraints could be lifted by tunneling from a region comparable in size to the Fermi wavelength ($\lambda_F$), where a large lateral momentum spread yields access to any momentum state in the other layer. Lifting of momentum constraints in tunneling is why the vertical, tunnel-resonant transistor mentioned earlier\cite{Simmons} fails to operate below a minimum device size. 

\begin{figure}[t]
\includegraphics[width=7.5cm]{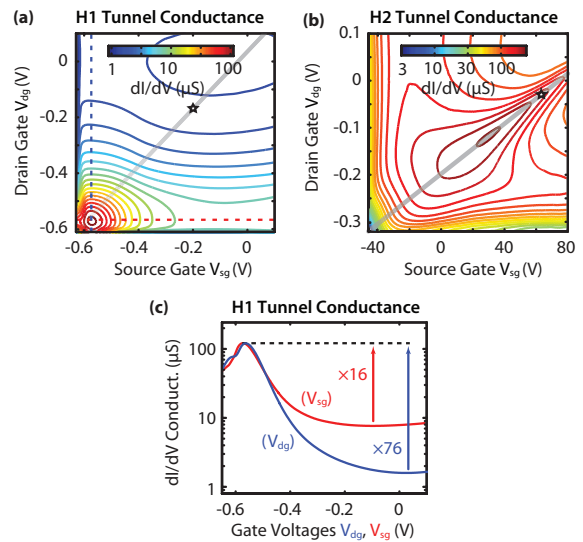} 
\caption{(a) Tunneling conductance as a function of source and drain gate voltages in the main heterostructure H1. Density-matched bilayer resonances (around gray line) are absent, but barrier-modulated tunnel increases along dotted lines are strong. (b) A different heterostructure H2 with resonant tunneling is shown for comparison. The axes of (a) and (b) are chosen so that both layers are depleted at the lower left corner, with a star added for reference where each layer has a carrier density of $2 \times 10^{11}$ cm$^{-2}$. (c) Cuts along dotted lines in (a) reveal the large tunnel increases. All data were taken at 4.2K.}
\label{Fig3}
\end{figure}

On the main heterostructure H1, the greatest \emph{relative} tunnel increase occurs when tuning either gate through the tunnel maximum at $(V_{sg},V_{dg})=$ (-0.57V, -0.57V). The corresponding source and drain gate sweeps, represented by the dotted lines in Fig 3a, show tunneling grow by a factor of 16 and 76, respectively (Fig 3c). Although the sweeps are not identical, their lineshapes are similar as expected from the symmetry in the heterostructure. Any quantitative difference likely comes from the mechanical and chemical processing that the source layer sees while the drain layer is protected face-down in epoxy.

The greater induced tunneling from the drain gate represents a change in effective conductivity from 9 nS to 600 nS/$\mu$m$^2$. The actual tunneling increase is probably larger than the measured conductance enhancement; as a layer is depleted, the growing tunnel conductance is eventually overcome by a low series sheet conductivity. This explains the apparent turnover of gate-modulated tunneling in Fig 3c. By independently measuring sheet resistance, we can estimate that the true tunnel conductance increases by at least another order of magnitude. The tunnel signal will not be obscured in this way if one combines smaller-area gates and higher-mobility heterostructures.

The robustness of wavefunction extension is demonstrated by its persistence even when a substantial bias is applied between layers. Here, we choose to focus on the effect of the drain gate rather than the source gate due to the drain gate's greater influence on tunneling. We measure the increase in tunneling from a nearly-depleted drain layer compared to an ungated drain layer, as a function of source-drain bias. For a positive source-drain bias, the relative gate-induced increase in current is unchanged for biases up to several mV, and some modulation is visible up to many tens of mV (Fig 4a). Low negative biases work similarly (Fig 4b), though wavefunction extension vanishes earlier at high negative bias (10 mV). This is probably because electrons begin to tunnel into the excited states of the drain regardless of its lowest subband energy (schematic: Fig 4b, inset). For positive bias, the drain subband is always the highest accessible level, so it affects tunneling for higher biases (Fig 4a, inset). 

\begin{figure}[t]
\includegraphics[width=7.5cm]{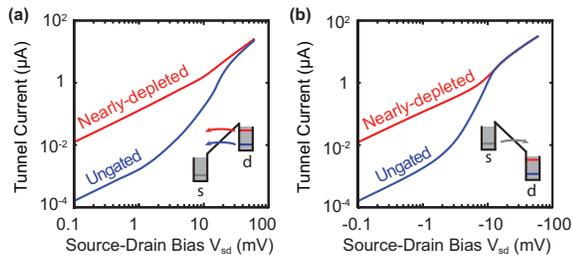} 
\caption{Current modulation by the drain gate as a function of (a) positive and (b) negative source-drain bias. The insets depict a possible explanation for the difference in modulation for large positive and negative biases.}
\label{Fig4}
\end{figure}

From Fig 4a, we can also calculate the transconductance, which peaks at 50 nS/$\mu$m$^2$ at $V_{sd} = 40$mV at 4.2K. Here, the units of transconductance are per area rather than per length, because of the unusual geometry of the transistor. For comparison, MOSFETs can achieve transconductances of 11-30 mS/$\mu$m with drives of around 70-200 mV at room temperature.\cite{Baker} To operate a WET at room temperature with higher transconductances, the effective barrier height needs to be made much larger while keeping the absolute barrier low. This can be accomplished by increasing $E_F$ of the source and drain layers. GaAs heterostructures are limited to Fermi energies of tens of meV. In contrast, graphene, a single atomic layer of graphitic carbon, has been gated to carrier densities\cite{graphene1} up to 3 $\times$ 10$^{13}$ cm$^{-2}$, which corresponds to $E_F$ = 0.9 eV.\cite{Electrolyte} This is 90 times larger than in our GaAs/AlGaAs heterostructure, with the additional advantage that graphene can be serially deposited\cite{Tutuc} and etched so that individual layers can be contacted without depletion gates. Furthermore, graphene is extremely thin, allowing for larger capacitances and transconductances with more closely-spaced gates.

Although graphene field-effect transistors have already been reported with large on-off current ratios at room temperature,\cite{Nanoribbon1, Nanoribbon2} such devices rely on nanoconstrictions to open bandgaps and have been predicted to have low yields for the near future because of difficult device fabrication.\cite{Geim} A graphene WET would not need precise lateral definition, and would instead depend on its more easily controlled vertical layer structure (Fig 5a). We have modeled (but not fabricated) a graphene WET containing two graphene sheets, doped to an easily-achievable $E_F=0.4$ eV (layer density of $6 \times 10^{12}$ cm$^{-2}$) and separated by a slightly $n$-doped silicon barrier. The barrier height is chosen so that there exists no excited interlayer subband below ten times room temperature thermal energy (250 meV). The graphene double layer is insulated on each side by a thin layer of high-k dielectric (HfO$_2$), and equal voltage is applied to top and back gates to match layer densities and energy-momentum dispersions. (Such a device should also have a high negative differential resistance\cite{Simmons} should the top and bottom gates be differently biased.)

\begin{figure}[b]
\includegraphics[width=7.5cm]{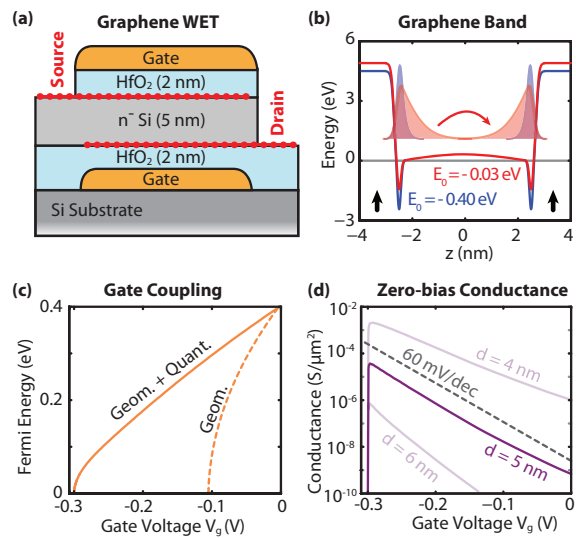} 
\caption{(a) Schematic of a graphene WET (two graphene layers marked by dotted red lines) and (b) plot of the conduction band with simultaneously varied subband wavefunctions (shaded). (c) Quantum capacitance, due to a discrete density of states, weakens the effect of the gate, but the device still is able to match 60 mV/decade over several decades (d).}
\label{Fig5}
\end{figure}

For a 5 nm-wide barrier and 2 nm top and back gate dielectrics, the layer wavefunctions significantly extend when the subbands are raised (Fig 5b). To estimate how gating might accomplish this, the density of states (DOS) of graphene must be considered. Unlike the constant DOS of the 2D GaAs heterostructure, graphene's DOS nominally drops to zero as it is depleted due to its linear dispersion. This introduces problems of quantum capacitance near zero DOS,\cite{GeimCap,Joey} which weakens the effect of a gate on $E_F$ as compared to that expected from the conventional geometric capacitance (Fig 5c). (This issue could potentially be alleviated by using bilayer graphene, which has a constant DOS near the K point.) Despite the shrinking DOSs, the tunnel current still increases with negative gate bias, almost to depletion, due to the exponentially increasing tunnel coupling (Fig 5d).

We have chosen the device parameters such that tunnel modulation is at the room-temperature thermal limit of 60 mV/decade over five decades. For a source-drain bias of 100 mV, where we have offset tunnel gate voltages to account for the bias-induced dispersion mismatch, we find that we should be able to obtain transconductances on the order of 10 $\mu$S/$\mu$m$^2$ over a 0.3V gate range. (Note again that transconductance scales with channel area rather than channel width, due to the geometry of the transistor.) If the tunnel barrier is made thicker (6 nm vs. 5 nm) the subthreshold slope can beat the thermal limit, provided phonon-assisted processes do not dominate, at the cost of somewhat reduced maximum conductance. We neglect the effect of spatial inhomogeneities in density,\cite{Yacoby,Crommie,DasSarma} because the fluctuations are on the order of the thermal energy and because the disorder that gives rise to these fluctuations continues to be reduced with improvements in fabrication technology.\cite{GrapheneBN} 

As mentioned previously, the original GaAs/AlGaAs WET is not practical for room-temperature operation, but does have an attractive potential application at low temperature. Using a scanned gate rather than a lithographically-patterned gate on a WET structure would allow local tunneling into complex, spatially-organized electron phases that sit at buried interfaces and are otherwise locally inaccessible. In this context, we have recently found empirically that the source and drain layers almost completely screen the effect of their respective gates on the opposing 2D layers, that non-equilibrium spectroscopy can be performed, and that tunneling spatial resolution should be on order of $\lambda_F$.\cite{VSTM} 

In summary, we have used a bilayer GaAs/AlGaAs heterostructure to demonstrate the soundness of the simple WET principle. The behavior of the tunnel modulation as a function of gate voltage and bias is understood qualitatively, and should permit the creation of WETs with useful specifications, using other materials. Furthermore, the GaAs/AlGaAs heterostructure as grown could serve as a tool for probing interesting physics. Again, although the WET design is not yet fully competitive with conventional transistors, wavefunction extension combined with new materials and fabrication techniques could lead to a new class of quantum transistors based on vertical transport in heterostructures.

We thank C.X. Liu for theoretical discussions. This work is supported by DOE-BES, DMS\&E at SLAC (DE-AC02-76SF00515), with the original concept developed under the Center for Probing the Nanoscale (NSF NSEC Grant No. 0425897) and a Mel Schwartz Fellowship from the Stanford Physics Department. This work was performed, in part, at the Center for Integrated Nanotechnologies, a DOE-BES user facility at Sandia National Labs (DE-AC04-94AL85000). The work at Princeton was partially funded by the Gordon and Betty Moore Foundation as well as the National Science Foundation MRSEC Program through the Princeton Center for Complex Materials (DMR-0819860). A.S. acknowledges support from  NSF, and M.P. from the Hertz Foundation, NSF, and Stanford. D.G.-G. recognizes support from the David and Lucile Packard Foundation

\end{document}